\documentclass{article}
\usepackage{amssymb}

\newcommand{\text}{\hbox}
\newcommand{\U}[1]{}

\begin{document}

\title{A classical model for a photodetector in the presence of electromagnetic vacuum
fluctuations}
\author{Trevor W. Marshall\\School of Mathematics,\\
University of Manchester,\\ Manchester M13 9PL, UK
\and Emilio Santos\\Departamento de 
F\'{i}sica,\\ Universidad de Cantabria,\\ Santander, Spain}


\maketitle

\abstract{
The main argument against the reality of the electromagnetic vacuum
fluctuations is that they do not activate photon detectors. In order to meet
this objection we propose a classical model of a photodetector which, in the
simple case of a light signal with constant intensity, gives a counting
rate which is a non-linear function of the intensity. For sufficiently large 
signal intensity,the counting rate is
proportional to the intensity, in agreement with the standard quantum
result, but there is a
dark rate when the signal intensity is low.

}

\section{\protect\smallskip Introduction}

The aim of this paper is to present a model of a photon counter resting upon
classical electromagnetic (Maxwell's) theory, but assuming that the whole
space is filled with a random radiation field (zeropoint field, ZPF). The
ZPF is assumed to have an average energy $\frac{1}{2}h\nu $ per normal mode, 
$\nu $ being the frequency and $h$ Planck's constant. Thus the ZPF
corresponds precisely to the vacuum fluctuations of quantum electrodynamics,
taken as a real field, and light signals would be radiation above that
``sea''\ of ZPF. The reader may ask what is the use of the model which we
propose because, according to the standard wisdom, classical Maxwell's
theory has been superseded by quantum electrodynamics and there exists
already a well established quantum theory of photon detection (see, e.g. the
book of Mandel\cite{Mandel}). In order to answer the question the present
section will be attempt to explain the relevance of our model.

More than eighty years have elapsed since the discovery of quantum
mechanics, but a vivid discussion still exists about its interpretation. The
various interpretations may be roughly classified in two groups which could
be called realistic and pragmatic. According to the pragmatic view, physics
just consists of a set of rules for the prediction of the results of
experiments. It is assumed that the question whether these rules lead to a
picture of the world does not belong to the domain of physics but rather to
the realm of philosophy. This view was supported by Niels Bohr and gave rise
to the \textquotedblleft Copenhagen interpretation\textquotedblright , which
has been accepted by the mainstream of the scientific community until
recently. In contrast, the realistic view claims that the essential aim of
natural sciences, physics in particular, is to provide knowledge about the
material world, the possibility of predicting the results of experiments
being a by-product. Of course it is an important by-product, because on it
rests the criterion for the validity of the theory, that is the agreement of
the predictions with the empirical results. The realistic view was
maintained by several of the founding fathers, in particular Albert
Einstein, and we strongly support it. Actually the realistic interpretation
follows the tradition of all sciences, not only physics, until the advent of
quantum mechanics. If the Copenhagen interpretation has had such a big
success it is because people have failed to find a general and coherent
realistic interpretation of quantum mechanics.

The difficulty for a realistic interpretation of quantum mechanics is that
it seems incompatible with the quantum formalism, at least if realism is
combined with the demand of locality, an incompatibility proved by John Bell
more than forty years ago\cite{bell}. During the eighty years of life of
quantum mechanics very many of its predictions have been tested and the
agreement between theory and experiments has been spectacular, especially in
quantum electrodynamics. Thus it is not strange that most people are so fond
of quantum mechanics that Bell's theorem has been taken as a proof of the
impossibility of a local realistic theory of the world, even without the
need of any additional empirical support. This prejudice has produced a bias
in the analysis of performed experiments, leading to the current wisdom that
local realism has been empirically refuted, modulo a few irrelevant loopholes%
\cite{laloe}. However, the truth is that all experiments performed till now
have had results compatible with local realism and the alleged empirical
disproof rests upon some unjustified extrapolations\cite{santosfp}. Of
course quantum mechanics has not been violated, so that local realism and
quantum mechanics are compatible for all experiments performed till now.
Therefore it is an open question whether they are complatible for all
experiments which can be actually performed. If this is true, then some of
the assumptions needed to prove Bell's theorem are flawed. This means that
either the quantum formalism or, most probably, the quantum measurement
theory should be modified.

The goal of devising a modification or alternative to quantum mechanics
compatible with local realism (and the experimients) is not easy. A possible
procedure is to begin with some theory which: a) provides from the start a
clear picture of the natural world, and b) allows an interpretation of the
experiments within some restricted domain. The said theory should be
generalized later in order to cover a wider domain. A theory of this type is 
\textit{stochastic electrodynamics (SED)}\cite{luissed} whose basic idea is
that quantum behaviour is related to the existence of a real ZPF.

The existence of vacuum fluctuations is a straightforward consequence of
field quantization \cite{milonni}. In addition, quantum vacuum fluctuations
have consequences which have been tested empirically. For instance, the
vacuum fluctuations of the electromagnetic field give rise to the main part
of the Lamb shift \cite{lamb} and to the Casimir effect \cite{cm}. Thus
Planck's radiation law should be written in the form 
\begin{equation}
\rho \left( \omega ,T\right) =\frac{\omega ^{2}}{\pi ^{2}c^{3}}\left[ \frac{%
\hbar \omega }{\exp \left( \hbar \omega /k_{B}T\right) -1}+\frac{1}{2}\hbar
\omega \right] =\frac{\hbar \omega ^{2}}{\pi ^{2}c^{3}}\coth \frac{\hbar
\omega }{k_{B}T}\quad ,  \label{planck}
\end{equation}
where the second term represents the ZPF. That the thermal spectrum contains
an $\omega ^{3}$ term has been proved by experiments measuring current
fluctuations in circuits with inductance at low temperature \cite{khc}. Of
course, the term is ultraviolet divergent so that some cutoff should be
assumed, possibly at about the Compton wavelength, where the fluctuations of
the Dirac electron-positron sea become important. The ZPF of SED is
identical with these quantum vacuum fluctuations, taken as a real stochastic
field.

The standard wisdom is that the vacuum fluctuations cannot be interpreted as
a \textit{real }random electromagnetic field because they do not activate
photodetectors in the absence of signals. There is also a gravitational
problem because, if the quantum vacuum fluctuations are at the origin of the
cosmological constant, as is usually assumed, that constant should be many
orders of magnitude larger than the observed value, but we shall not be
concerned with gravitational effects in this paper. A common explanation of
the fact that the zeropoint field (ZPF)\ does not activate photodetectors is
to say that the ZPF is not \textit{real}, but \textit{virtual}. In our
opinion replacing a word, real, by another one, virtual, with a less clear
meaning is not a good solution. In the present article we shall prove that
the behaviour of photodetectors can be explained without renouncing the
reality of the ZPF. The proof goes via constructing an explicit model a of
detector, producing a counting rate proportional to the intensity of the
signal, that is also able to subtract efficiently the ZPF. The mere
existence of the model proves that the said objection to the ZPF reality is
untenable. Such a proof is the first purpose of the present paper.

In addition we think that a classical model of detection may add to
understand the origin of an open problem lasting for more than 30 years,
namely the existence of a \textit{detection loophole. }In fact, as is well
known all experiments aimed at a discrimination between quantum mechanics
and local realism, via the test of Bell's inequalities using optical
photons, suffer from a loophole due to the lack of efficient detectors. The
standard wisdom is that the difficulty in manufacturing efficient photon
counters is a minor technical problem. However the persistence of the
difficulty for so many years may indicate that the problem is a fundamental
one. The comparison of the classical model that we propose with the quantum
detection theory may throw light on the problem, and this is another purpose
of the present paper.

\section{Stochastic properties of the zeropoint field}

Now we shall derive some relevant properties of the ZPF in free space, that
is far from any material body. The ZPF is characterized by the electric
field \textbf{E}(\textbf{r},t), and the magnetic field \textbf{B}(\textbf{r}%
,t). The stochastic properties of the field may be summarized saying that it
is Gaussian, so that the mean and the correlation functions are sufficient
to characterize all stochastic properties. The mean is zero and the
correlation functions of the field components are\cite{marsh}\cite{boyer} 
\begin{eqnarray}
\left\langle E_{i}(\mathbf{r},t)E_{j}(\mathbf{r}^{\prime },t^{\prime
})\right\rangle  &=&\int d^{3}k\left( \delta _{ij}-\frac{k_{i}k_{j}}{k^{2}}%
\right) \frac{\U{127}h{\hskip-.2em}\llap{\protect\rule[1.1ex]{.325em}{.1ex}}{%
\hskip.2em}\omega }{4\pi ^{2}}\cos \left[ \mathbf{k}\cdot\left( \mathbf{r-r}%
^{\prime }\right) -\omega \left( t-t^{\prime }\right) \right] ,  \nonumber \\
\left\langle B_{i}(\mathbf{r},t)B_{j}(\mathbf{r}^{\prime },t^{\prime
})\right\rangle  &=&\left\langle E_{i}(\mathbf{r},t)E_{j}(\mathbf{r}^{\prime
},t^{\prime })\right\rangle ,  \nonumber \\
\left\langle E_{i}(\mathbf{r},t)B_{j}(\mathbf{r}^{\prime },t^{\prime
})\right\rangle  &=&\int d^{3}k\varepsilon _{ijl}\frac{k_{l}}{k}\frac{\U{127}%
h{\hskip-.2em}\llap{\protect\rule[1.1ex]{.325em}{.1ex}}{\hskip.2em}\omega }{%
4\pi ^{2}}\cos \left[ \mathbf{k}\cdot
\left( \mathbf{r-r}^{\prime }\right)
-\omega \left( t-t^{\prime }\right) \right] ,  \label{boyer}
\end{eqnarray}
where k = $\left| \mathbf{k}\right| =\omega /c.$ The integrals in these
expressions do not converge but a natural cut-off in $\omega $ appears as
explained in the following. Our aim is to make a photodetection model and we
may assume that any detection event takes place essentially at a single
atom. Thus the quantity of interest for us will be the time autocorrelation
of the averages of the fields over the volume of the atom. We define the
averaged quantities 
\begin{equation}
E_{j}(t)\equiv \int d^{3}rE_{j}(\mathbf{r},t)\rho \left( r\right)
,\;B_{j}(t)\equiv \int d^{3}rB_{j}(\mathbf{r},t)\rho \left( r\right) ,
\label{ave}
\end{equation}
where no confusion should arise by the use of the same label for the field
at a point and the averaged quantity. The function $\rho \left( r\right) $
represents a normalized effective electron density in the atom, which I
assume spherically symmetrical. From eqs.$\left( \ref{boyer}\right) $ it is
straightforward to get the time correlations of the averaged quantities and
we get 
\begin{equation}
\left\langle E_{j}(t)E_{k}(t^{\prime })\right\rangle =\left\langle
B_{j}(t)B_{k}(t^{\prime })\right\rangle =\delta _{jk}F(t-t^{\prime
}),\;\left\langle E_{j}(t)B_{k}(t^{\prime })\right\rangle =0,  \label{em}
\end{equation}
where 
\begin{equation}
F(\tau )=\frac{32\pi \U{127}h{\hskip-.2em}\llap{\protect%
\rule[1.1ex]{.325em}{.1ex}}{\hskip.2em}}{3c}\int_{0}^{\infty }\omega \cos
\left( \omega \tau \right) d\omega \left[ \int_{0}^{\infty }\sin \left(
\omega r/c\right) \rho \left( r\right) rdr\right] ^{2}.  \label{F}
\end{equation}
For any reasonable density, $\rho \left( r\right) $, the function $F(\tau
)=F\left( -\tau \right) $ decreases slowly for $\left| \tau \right| \lesssim
a/c$ and rapidly for higher $\tau $, becoming negligible for $\left| \tau
\right| >>a/c,$ $c$ being the velocity of light and $a$ the atomic radius.

For our detection model the most relevant quantity is the radiation
intensity represented by the Poynting vector, 
\[
\mathbf{S(}t)=\frac{c}{4\pi }\mathbf{E}(t)\times \mathbf{B}(t), 
\]
where $\mathbf{E}(t)$ and $\mathbf{B}(t)$ are the averages over the volume
of the atom defined in $\left( \ref{ave}\right) $. The mean of the Poynting
vector is zero and the correlation functions of its components may be
derived from $\left( \ref{em}\right) $ to be 
\begin{equation}
\left\langle S_{j}\mathbf{(}t)S_{k}\mathbf{(}t^{\prime })\right\rangle =%
\frac{c^{2}}{8\pi ^{2}}\delta _{jk}F\left( t^{\prime }-t\right) ^{2}\simeq
\sigma ^{2}\delta _{jk}\delta \left( t^{\prime }-t\right) ,  \label{point}
\end{equation}
where $\delta _{jk}$ is Kronecker\'{}s delta, $\delta \left( t^{\prime
}-t\right) $ is Dirac\'{}s delta and $\sigma $ is a constant of the order of 
$10^{-3}\U{127}c^{3/2}a^{-7/2}.$ The second equality involves approximating
every component of the Poynting vector by a white noise, which is plausible
provided we study detection of light signals whose coherence time, $\tau ,$
is much larger than the time interval where $F(\tau )^{2}$ is not
negligible, that is $\tau >>a/c.$ These properties characterize the
components of the Poynting vector as three stationary stochastic processes,
statistically independent and having the autocorrelation of a white noise.

Actually we are interested in the ZPF within a photon detector and the
question is whether we may use there the statistical properties $\left( \ref
{point}\right) $ of free space. Our answer is in the affirmative because we
should assume that the ZPF permeates everything. Indeed in SED it is
supposed that the vacuum fluctuations of the electromagnetic field - and
possibly other fields also - is the cause of the random position of the
electron inside the atom. In the standard interpretation of quantum
mechanics it is also assumed that the vacuum fields fluctuate even inside
material bodies, with an average energy $\frac{1}{2}h\nu $ per normal mode,
although the modes are modified by the presence of matter (this gives rise,
for instance, to the Casimir effect and the Lamb shift as mentioned above).
However the radiation modes which are most modified by the presence of
matter are those of low frequency, much lower than the ones most strongly
involved in $F(\tau ),$ eq.$\left( \ref{F}\right) .$ Thus we may use the
free space statistical properties $\left( \ref{point}\right) $ in our model.

We shall study the situation where we have a determinate light signal
superimposed on the ZPF. The statistical independence of signal and random
fields leads to (compare with $\left( \ref{point}\right) )$%
\begin{equation}
\left\langle S_{j}\mathbf{(}t)\right\rangle =\delta _{j3}I_{s},\left\langle
S_{j}\mathbf{(}t)S_{k}\mathbf{(}t^{\prime })\right\rangle =\delta
_{j3}\delta _{k3}I_{s}^{2}+\sigma ^{2}\delta _{jk}\delta \left( t^{\prime
}-t\right) ,  \label{poyntin}
\end{equation}
where $S_{3}$ is the component of the Poynting vector in the direction of
the beam and $I_{s}$ the signal intensity. Eq.$\left( \ref{poyntin}\right) $
will be the basis of our subsequent study but we should bear in mind that
the approximations involved may be too crude for some applications. Possible
improvements will be considered elsewhere.

\section{Detection model}

Several models of photon counter have been proposed in the context of SED,
resting upon the idea that there exists a \textquotedblleft detection
time\textquotedblright , T, independent of the light intensity and such that
the probability of a count depends on the radiation, including the ZPF,
which enters the detector during the time $T$\cite{crs}. Those models,
however, are not compatible with empirical evidence\cite{s}. Instead of
fixing the detection time $T$, here we shall assume that every atom
accumulates the energy and momentum of the radiation arriving at it and a
count is produced whenever that energy surpasses some threshold. If $\mathbf{%
S}(t)$ is the Poynting vector of the radiation arriving at the detecting
atom at time $t$, the accumulated energy at time $T$ will be 
\begin{equation}
E(T)=A\left\vert \int_{0}^{T}\mathbf{S}(t)dt\right\vert ,  \label{energy}
\end{equation}
where $A$ is the effective cross section of the atom. In the following we
shall put $A=1$ in calculating the detection rate and multiply times $A$ at
the end.

The essential assumption of our model is that \textit{a detection event is
produced at a time }$T$\textit{, after the previous count, when }$T$\textit{%
\ is such that } 
\begin{equation}
E\left( T\right) =\left\vert \mathbf{E}\left( T\right) \right\vert =E_{m},\;%
\mathbf{E}(t)\equiv \int_{0}^{t}\mathbf{S}(t^{\prime })dt^{\prime }\quad ,
\label{model}
\end{equation}
\textit{and }$E$\textit{\ is a parameter characteristic of the detector. }

The calculations resting upon the model $\left( \ref{model}\right) $ are
involved due to the fluctuations of the ZPF and the signal. Indeed
constructing a detailed detection model on the basis of that equation would
require using the theory of \textquotedblleft first passage
time\textquotedblright\ for the (vector) stochastic process $\mathbf{E}(t),$
which has a finite, nonzero, correlation time. The problem is dramatically
simplified if we assume that every cartesian component of $\mathbf{E}(t)$ is
a white noise (having a null correlation time) superimposed on a
deterministic signal with constant intensity $I_{s}$, as in eq.$\left( \ref
{poyntin}\right) $. More specifically, we assume that every component of the
stochastic process $\mathbf{E}(t)$ is a Wiener (Brownian motion) process.
The calculation of the first-passage time is now straightforward. We must
begin solving the diffusion equation 
\begin{equation}
\frac{\partial \rho }{\partial t}=\frac{1}{2}\sigma ^{2}\left( \frac{%
\partial ^{2}\rho }{\partial E_{x}^{2}}+\frac{\partial ^{2}\rho }{\partial
E_{y}^{2}}+\frac{\partial ^{2}\rho }{\partial E_{z}^{2}}\right) -I_{s}\frac{%
\partial \rho }{\partial E_{z}},\;\rho =\rho \left( \mathbf{E},t\right) ,
\label{ode}
\end{equation}
with the initial condition 
\begin{equation}
\rho \left( \mathbf{E},0\right) =\delta ^{3}\left( \mathbf{E}\right) \equiv
\delta \left( E_{x}\right) \delta \left( E_{y}\right) \delta \left(
E_{z}\right) ,  \label{initial}
\end{equation}
and an absorbing barrier at $\left\vert \mathbf{E}\right\vert =E_{m},$ that
is the boundary condition 
\begin{equation}
\rho \left( \mathbf{E},t\right) =0\text{ for }\left\vert \mathbf{E}%
\right\vert =E_{m}.  \label{boundary}
\end{equation}

The solution of eq.$\left( \ref{ode}\right) $ with these conditions is
cumbersome due to the fact that the symmetry properties of the boundary
condition, eq.$\left( \ref{boundary}\right) ,$ are different from those of
the diffusion eq.$\left( \ref{ode}\right) $ itself. In order to simplify the
calculations, whithout changing qualitatively the model, we substitute for
eq.$\left( \ref{boundary}\right) $ the following 
\begin{equation}
\rho \left( E_{x},E_{y},E_{z},t\right) =0\text{ for either }E_{x}=\pm E_{m}%
\text{ or }E_{y}=\pm E_{m}\text{ or }E_{z}=\pm E_{m}.  \label{bound1}
\end{equation}
This means replacing a sphere in the 3D space of the vectors $\mathbf{E}$ by
a cube. Thus our problem may be reduced to solving 3 one-dimensional
equations by introducing the functions $f_{x},f_{y},f_{z}$ such that 
\[
\rho \left( E_{x},E_{y},E_{z},t\right) =f_{x}\left( E_{x},t\right)
f_{y}\left( E_{y},t\right) f_{z}\left( E_{z},t\right) . 
\]
Indeed the solution of eq.$\left( \ref{ode}\right) $ with conditions eqs.$%
\left( \ref{initial}\right) $ and $\left( \ref{bound1}\right) $ may be
obtained from the solution of the three ordinary differential equations 
\begin{equation}
\frac{\partial f_{x}}{\partial t}=\frac{1}{2}\sigma ^{2}\frac{\partial
^{2}f_{x}}{\partial E_{x}^{2}},\frac{\partial f_{y}}{\partial t}=\frac{1}{2}%
\sigma ^{2}\frac{\partial ^{2}f_{y}}{\partial E_{y}^{2}},\frac{\partial f_{z}%
}{\partial t}=\frac{1}{2}\sigma ^{2}\frac{\partial ^{2}f_{z}}{\partial
E_{z}^{2}}-I_{s}\frac{\partial f_{z}}{\partial E_{z}},  \label{odes}
\end{equation}
with initial and boundary conditions, respectively, 
\[
f_{x}\left( E_{x},0\right) =\delta \left( E_{x}\right) ,\;f_{x}\left(
E_{m},t\right) =f_{x}\left( -E_{m},t\right) =0, 
\]
and similar for $f_{y}$ and $f_{z}$. The appropriate solutions of eqs.$%
\left( \ref{odes}\right) $ may be obtained by the method of the images and
we get (we shall put $\sigma ^{2}=2$ in the calculation and restore $\sigma $
in the final result$)$ 
\begin{equation}
f_{z}\left( E_{z},t\right) =\frac{1}{\sqrt{4\pi t}}\exp \left[ \frac{1}{2}%
E_{z}I_{s}\right] \sum_{n=-\infty }^{\infty }\left( -1\right) ^{n}\exp
\left[ -\frac{\left( E_{z}+2nE_{m}\right) ^{2}}{4t}-\frac{I_{s}^{2}t}{4}%
\right] ,  \label{roet}
\end{equation}
and similar expressions, except replacing $I_{s}$ by $0$, for $f_{x}\left(
E_{x},t\right) $ and $f_{y}\left( E_{y},t\right) .$

Now our model predicts that, if we have a detection event at time $t\ =0$,
the probability that the next detection event takes place before time $t$ is 
\begin{equation}
P(t)=1-\int_{-E_{m}}^{E_{m}}f_{x}\left( E_{x},t\right)
dE_{x}\int_{-E_{m}}^{E_{m}}f_{y}\left( E_{y},t\right)
dE_{y}\int_{-E_{m}}^{E_{m}}f_{z}\left( E_{z},t\right) dE_{z}.  \label{P}
\end{equation}
Our aim is to calculate the detection rate, which is the inverse of the mean
first passage time, that is 
\begin{equation}
<t>=\int_{0}^{\infty }t\frac{dP(t)}{dt}dt.  \label{mftp}
\end{equation}
The proof that this average gives the inverse of the detection rate is as
follows. We consider that the detector is active during a very long time
interval. Within it we will have a large number of detection events on each
atom. Let us assume, for the sake of clarity, that the time intervals
between two detection events form a discrete sequence t$_{1}$, t$_{2}$, ...t$%
_{j}$,... If we have N$_{j}$ time intervals of duration t$_{j}$ then the
detection rate will be 
\begin{equation}
R=\frac{\sum N_{j}}{\sum N_{j}t_{j}}=\frac{1}{\sum P_{j}t_{j}}=\frac{1}{<t>},
\label{R}
\end{equation}
where P$_{j}$ is the probability that a time interval between two detection
events has duration $t_{j}$. If we pass to the continuous, we shall replace
the summation by an integral, giving a rate $R$ equal to the inverse of $<t>$%
, which completes the proof. The counting rate in a macroscopic detector
consisting of many atoms may be obtained from the predicted rate in one atom
but this will not be made in the present article.

According to eqs.$\left( \ref{P}\right) $ and $\left( \ref{mftp}\right) $the
mean ``first passage time'' is given by

\begin{eqnarray}
&<&t>=\int_{0}^{\infty }dtLK^{2},\;L\equiv \int_{-E_{m}}^{E_{m}}f_{z}\left(
E_{z},t\right) dE_{z}  \label{t} \\
K &\equiv &\int_{-E_{m}}^{E_{m}}f_{x}\left( E_{x},t\right) dE_{x}\equiv
\int_{-E_{m}}^{E_{m}}f_{y}\left( E_{y},t\right) dE_{y}.  \nonumber
\end{eqnarray}
Hence the calculation is straighforward putting eqs.$\left( \ref{roet}%
\right) $ in eq.$\left( \ref{t}\right) $ and performing the integrals. But
before proceeding at this calculation we will work a one-dimensional model
consisting of the replacement of $K(t)$ by unity, which corresponds to
substituting a layer for the atom. As we shall see that model gives
predictions not too different from the three-dimensional one and it has the
advantage of leading to a simple analytical expression for the detection
rate. From eqs.$\left( \ref{roet}\right) $ and $\left( \ref{t}\right) $ we
obtain, making the time integral first, 
\begin{eqnarray}
&<&t>==\int_{0}^{\infty }dtL=\int_{-Em}^{Em}dE\;\frac{1}{\sqrt{4\pi }}\exp
\left[ \frac{1}{2}EI_{s}\right] \sum_{n=-\infty }^{\infty }\left( -1\right)
^{n}J_{n},  \label{L} \\
J_{n} &=&\int_{0}^{\infty }dt\frac{1}{\sqrt{t}}\exp \left[ -\frac{\left(
E+2nE_{m}\right) ^{2}}{4t}-\frac{I_{s}^{2}t}{4}\right] =\frac{\sqrt{4\pi }}{%
I_{s}}\exp \left[ -\frac{1}{2}I_{s}\left\vert E+2nE_{m}\right\vert \right] ,
\nonumber
\end{eqnarray}
where the integral has been taken from Gradshteyn\cite{grad} . In order to
perform the sum in $n$ it is convenient to separate terms with $n=0,$ $n>0$
and $n<0$, giving 
\begin{eqnarray}
\text{Sum} &=&\exp \left[ -\frac{1}{2}I_{s}\left\vert E\right\vert \right]
+\sum_{n=1}^{\infty }\left( -1\right) ^{n}\left\{ \exp \left[ -\frac{1}{2}%
I_{s}(2nE_{m}+E)\right] +\exp \left[ -\frac{1}{2}I_{s}(2nE_{m}-E)\right]
\right\}  \nonumber \\
&=&\exp \left[ -\frac{1}{2}I_{s}\left\vert E\right\vert \right] -2\left(
1+\exp \left[ I_{s}E_{m}\right] \right) ^{-1}\cosh \left[ \frac{1}{2}%
I_{s}E\right] .\smallskip  \label{L1}
\end{eqnarray}
The integral in $E$ is now easy and we obtain 
\begin{equation}
<t>=\frac{E_{m}}{I_{s}}\tanh \left( \frac{1}{2}I_{s}E_{m}\right)  \label{R0}
\end{equation}
and, restoring the parameter $\sigma ,$%
\begin{equation}
R=\frac{1}{<t>}=\frac{I_{s}}{E_{m}}\coth \left( \frac{I_{s}E_{m}}{\sigma ^{2}%
}\right) ,  \label{R1}
\end{equation}
which has the limits 
\begin{equation}
R\simeq \frac{I_{s}}{E_{m}}\text{ }(1+2\exp \left( -2\frac{I_{s}E_{m}}{%
\sigma ^{2}}\right) )\text{ }for\text{ }I_{s}>>\frac{\sigma ^{2}}{E_{m}},%
\text{ }R\simeq \frac{\sigma ^{2}}{E_{m}^{2}}\text{ }for\text{ }I_{s}<<\frac{%
\sigma ^{2}}{E_{m}}.  \label{R4}
\end{equation}

In order to solve the 3D model we begin by calculating 
\begin{eqnarray}
K\left( t\right) &=&\frac{1}{\sqrt{4\pi t}}\int_{-E_{m}}^{E_{m}}dE_{x}%
\sum_{n=-\infty }^{\infty }\left( -1\right) ^{n}\exp \left[ -\frac{\left(
E_{x}+2nE_{m}\right) ^{2}}{4t}\right]  \nonumber \\
&=&\frac{1}{\sqrt{4\pi t}}\sum_{n=-\infty }^{\infty }\int_{\left(
2n-1\right) E_{m}}^{\left( 2n+1\right) E_{m}}\left( -1\right) ^{n}\exp
\left( -\frac{s^{2}}{4t}\right) ds,  \label{K}
\end{eqnarray}
where the last expression is obtained by means of a trivial change of
variable. We see that $K(t)$ is the integral of a Gaussian times a function, 
$F(s)$, taking alternately the values $+1$ and $-1$ in intervals of length $%
2E_{m}$. We will perform the integral using the Fourier expansion of this
function, that is 
\[
F(s)=\sum_{k=0}^{\infty }\frac{4}{\left( 2k+1\right) \pi }\left( -1\right)
^{k}\cos \left( \frac{\left( 2k+1\right) \pi s}{2E_{m}}\right) , 
\]
whence we get 
\begin{eqnarray}
K\left( t\right) &=&\frac{4}{\pi \sqrt{4\pi t}}\sum_{k=0}^{\infty }\frac{%
\left( -1\right) ^{k}}{2k+1}\int_{-\infty }^{\infty }\cos \left( \frac{%
\left( 2k+1\right) \pi s}{2E_{m}}\right) \exp \left( -\frac{s^{2}}{4t}%
\right) ds=\sum_{k=0}^{\infty }K_{k}\left( t\right) ,  \nonumber \\
K_{k}\left( t\right) &=&\frac{4}{\pi }\frac{\left( -1\right) ^{k}}{2k+1}\exp
\left( -\frac{\left( 2k+1\right) ^{2}\pi ^{2}t}{4E_{m}^{2}}\right) .
\label{K1}
\end{eqnarray}
After that we have, from eqs.$\left( \ref{K1}\right) $and $\left( \ref{t}%
\right) $, 
\begin{equation}
<t>=\int_{0}^{\infty }\frac{dt}{\sqrt{4\pi t}}K\left( t\right)
^{2}\int_{-E_{m}}^{E_{m}}dE\exp \left[ \frac{1}{2}EI_{s}\right]
\sum_{n=-\infty }^{\infty }\left( -1\right) ^{n}\exp \left[ -\frac{\left(
E+2nE_{m}\right) ^{2}}{4t}-\frac{I_{s}^{2}t}{4}\right] .  \label{t1}
\end{equation}
Making the time integral first we get (compare with eq.$\left( \ref{L}%
\right) )$%
\begin{eqnarray}
&<&t>=\frac{16}{\pi ^{2}}\sum_{k=0}^{\infty }\sum_{l=0}^{\infty
}\int_{-Em}^{Em}dE\;\frac{1}{\sqrt{4\pi }}\exp \left[ \frac{1}{2}%
EI_{s}\right] \sum_{n=-\infty }^{\infty }\left( -1\right) ^{n+k+l}\frac{1}{%
\left( 2k+1\right) \left( 2l+1\right) }J_{nkl},  \nonumber \\
J_{nkl} &=&\int_{0}^{\infty }dt\frac{1}{\sqrt{t}}\exp \left[ -\frac{\left(
E+2nE_{m}\right) ^{2}}{4t}-\frac{I_{kl}^{2}t}{4}\right] =\frac{\sqrt{4\pi }}{%
I_{kl}}\exp \left[ -\frac{1}{2}I_{kl}\left\vert E+2nE_{m}\right\vert \right]
,  \label{t2}
\end{eqnarray}
where 
\begin{equation}
I_{kl}=\sqrt{I_{s}^{2}+\frac{[\left( 2k+1\right) ^{2}+\left( 2l+1\right)
^{2}]\pi ^{2}}{E_{m}^{2}}}.  \label{I}
\end{equation}
Hence 
\[
<t>=\sum_{k=0}^{\infty }\sum_{l=0}^{\infty }\frac{16}{\pi ^{2}I_{kl}}%
\int_{-Em}^{Em}dE\exp \left[ \frac{1}{2}I_{s}E\right] \;\sum_{n=-\infty
}^{\infty }\left( -1\right) ^{n}\exp \left[ -\frac{1}{2}I_{kl}\left\vert
E+2nE_{m}\right\vert \right] . 
\]
The sum in $n$ is similar to that in $\left( \ref{L1}\right) $ with $I_{kl}$
substituted for $I_{s}$ , the integrals in $E$ are straightforward and we
obtain 
\begin{eqnarray}
&<&t>=\frac{64}{\pi ^{2}}\sum_{k=0}^{\infty }\sum_{l=0}^{\infty }\frac{%
\left( -1\right) ^{k+l}}{[I_{kl}^{2}-I_{s}^{2}]\left( 2k+1\right) \left(
2l+1\right) }\left[ 1-\frac{\cosh \left( \frac{1}{2}I_{s}E_{m}\right) }{%
\cosh (\frac{1}{2}I_{kl}E_{m})}\right]  \nonumber \\
&=&\frac{128}{\pi ^{4}}\frac{E_{m}^{2}}{\sigma ^{2}}F\left( \frac{I_{s}E_{m}%
}{\sigma ^{2}}\right) ,  \nonumber \\
F\left( x\right) &\equiv &\sum_{k=0}^{\infty }\sum_{l=0}^{\infty }\frac{%
\left( -1\right) ^{k+l}}{[\left( 2k+1\right) ^{2}+\left( 2l+1\right)
^{2}]\left( 2k+1\right) \left( 2l+1\right) }G(x),  \nonumber \\
G(x) &=&1-\frac{\cosh x}{\cosh \sqrt{x^{2}+\frac{1}{4}\pi ^{2}\left[ \left(
2k+1\right) ^{2}+\left( 2l+1\right) ^{2}\right] }},  \label{t4}
\end{eqnarray}
where we have restored the parameter $\sigma $ (see eq.$\left( \ref{roet}%
\right) ).$ The function $F(x)$ is defined by an infinite series which
should be calculated numerically. Here we shall make only the calculation
for large and small values of $x$.

For $x>>1$ we may write (using the approximation $\cosh \alpha \simeq \frac{1%
}{2}\exp \alpha $ , valid for $\alpha >>1)$%
\begin{eqnarray}
G(x) &\simeq &1-\exp \left\{ x-\sqrt{x^{2}+\frac{1}{4}\pi ^{2}\left[ \left(
2k+1\right) ^{2}+\left( 2l+1\right) ^{2}\right] }\right\}   \nonumber \\
&\simeq &1-\exp \left\{ -\frac{\pi ^{2}\left[ \left( 2k+1\right) ^{2}+\left(
2l+1\right) ^{2}\right] }{8x}+O\left( \frac{1}{x^{3}}\right) \right\} .
\label{Gx}
\end{eqnarray}
Hence it is easy to get $F(x)$ for large $x$, that is 
\[
F\left( x\right) \sim \frac{\pi ^{2}}{8x}\sum_{k=0}^{\infty
}\sum_{l=0}^{\infty }\frac{\left( -1\right) ^{k+l}}{\left( 2k+1\right)
\left( 2l+1\right) }=\frac{\pi ^{2}}{8x}\left( \sum_{k=0}^{\infty }\frac{%
\left( -1\right) ^{k}}{\left( 2k+1\right) }\right) ^{2}=\frac{\pi ^{4}}{128x}%
,
\]
whence we obtain 
\[
<t>=\frac{E_{m}}{I_{s}}+O\left( \frac{\sigma ^{2}}{I_{s}^{2}}\right)
\rightarrow R\simeq \frac{I_{s}}{E_{m}},
\]
which is similar to the result of the one-dimensional model but the
deviation from the standard quantum result decreases as $1/I_{s}^{2}$
instead of exponentially (see eq.$\left( \ref{R4}\right) .)$

For $x=0$ we get, from eqs.$\left( \ref{t4}\right) $ and $\left( \ref{I}%
\right) ,$ 
\[
F(0)=\sum_{k=0}^{\infty }\sum_{l=0}^{\infty }\frac{\left( -1\right)
^{k+l}\left[ 1-1/\cosh \left( \frac{\pi }{2}\sqrt{\left( 2k+1\right)
^{2}+\left( 2l+1\right) ^{2}}\right) \right] }{\left( 2k+1\right) \left(
2l+1\right) \left[ \left( 2k+1\right) ^{2}+\left( 2l+1\right) ^{2}\right] }. 
\]
The series converges quickly and a numerical calculation gives 
\[
<t>\simeq 0.49\frac{E_{m}^{2}}{\sigma ^{2}}\rightarrow R\simeq 2.0\frac{%
\sigma ^{2}}{E_{m}^{2}}, 
\]
that is a rate at zero intensity about twice that in the one-dimensional
model, eq.$\left( \ref{R4}\right) .$

It is remarkable that we obtain a detection rate proportional to the signal
intensity, that is a perfect subtraction of the ZPF, when the intensity of
the signal is big, a result in agreement with the quantum mechanical
prediction. However for low, or nil, intensity signals there is some
counting rate, which we should interpret as a ``fundamental dark rate'' of
the detector.

The result may be generalized to the case where the signal intensity is not
a constant, but a known function of time. It would be enough to substitute $%
\int_{0}^{t}I_{s}(t^{\prime })dt^{\prime }$ for $I_{s}t$ in the above
equations, although the solution of the differential equation $\left( \ref
{ode}\right) $ would be more involved. More difficult would be to treat the
case where the signal itself fluctuates (with a correlation time of the
order of the inverse of the frequency bandwidth). We shall not study these
problems here.

We might also analyze coincidence counts in two detectors when the incoming
beams, with intensities $I_{1}(t)$ and $I_{2}(t)$ \textit{above the ZPF, }%
are correlated. The calculation would be straightforward, although lengthy.
We may assume that eq.$\left( \ref{R4}\right) $ is still valid for each
detector and the coincidence rate, with a time delay $\tau ,$ will be 
\begin{equation}
R_{12}\propto \left\langle I_{1}(t)I_{2}(t+\tau )\right\rangle ,
\label{coin}
\end{equation}
again in agreement with the quantum prediction. However the current
situation may not be that. In practice the signals may be stochastic and not
independent of the ZPF. The calculation in these conditions would be rather
involved.

\section{Discussion}

Our analysis shows that quantum vacuum fluctuations of the electromagnetic
field (or ZPF) may be efficiently subtracted by a\textit{\ }model which
assumes that the radiation is a classical (Maxwell) field including a
fluctuating ZPF, provided that the fluctuations of the signal have a large
enough correlation time in comparison with the effective correlation time of
the ZPF. This is usually the case in astronomical observations. In contrast,
in standard quantum optical experiments the fluctuations of the signal may
have a rather short correlation time. If the correlation time of the signal
does not fulfil the assumptions of the previous section, the presence of the
ZPF will probably give rise to deparatures from the standard quantum
predictions eqs.$\left( \ref{R1}\right) $ and $\left( \ref{coin}\right) $,
that is they will produce some nonidealities in the behaviour of optical
photon counters. This is specially important when it is necessary to measure
coincidence counting rates with short time windows, as is frequent in
quantum optical experiments (e.g. optical tests of Bell\'{}s inequality). If
this is the case, our approach may provide an explanation for the
difficulties of performing loophole-free tests of Bell\'{}s inequality using
optical photons.

I emphasize that, although our model is semiclassical, probably its main
properties might be reproduced by a rigorous quantum treatment. Furthermore
the difficulties for reaching an intuitive picture of how detectors subtract
the ZPF probably do not derive from quantum theory itself, but from the use
of approximations like first-order perturbation theory or taking the limit
of time $t\rightarrow \infty $ in calculating the probability of photon
absorption per unit time. Indeed I conjecture that excesive idealizations
might be at the origin of the difficulties for undersanding intuitively the
paradoxical aspects of quantum physics. Although simplifications are
extremely useful for calculations, they tend to obscure the physics.

\section{Conclusion}

We want to compare our rate prediction 
\begin{equation}
R_{m}=\frac{1}{<t>}=\frac{I_{s}}{E_{m}}\coth \left( \frac{I_{s}E_{m}}{\sigma
^{2}}\right) ,  \label{Rmodel}
\end{equation}
with the quantum prediction 
\begin{equation}
R_{q}=\eta KI_{s},  \label{Rquantum}
\end{equation}
where $\eta $ is the quantum efficiency and $K$ is a dimensional constant
taking into account the energy of one \textquotedblleft
photon\textquotedblright , the effective cross section of an atom and the
number of atoms in the detector. The two predictions are identical for high
intensity provided we identify 
\begin{equation}
E_{m}=\frac{1}{K\eta }.  \label{Em}
\end{equation}
But for low intensity and/or high efficiency there is a departure. For
instance the departure becomes more than 10\% (that is the \textquotedblleft
fundamental dark rate\textquotedblright\ is more than 10\% of the total
rate) when $I_{s}E_{m}/\sigma ^{2}<1.5.$ Using $\left( \ref{Em}\right) $
this gives 
\[
I_{s}<1.5K\sigma ^{2}\eta , 
\]
that is when the intensity is too low and/or the quantum efficiency too
high. As the tests of Bell's inequalities requires both high efficiency and
low intensity our model may explain the difficulty, or maybe impossibility
of such tests . A more quantitative result would require an estimate of $K$,
which is not easy.

\section{Appendix}

\subsection{Evaluation of F$\left( \tau \right) $ with $\rho \left( r\right)
=\left( 8\pi a^{3}\right) ^{-1}\exp \left( -r/a\right) $}

The density $\rho \left( r\right) $ is normalized in the sense $%
\int_{0}^{\infty }\rho \left( r\right) 4\pi r^{2}dr=1.$

We start calculating 
\[
\int_{0}^{\infty }\sin \left( \omega r/c\right) \rho \left( r\right) rdr=%
\frac{c^{3}\omega }{4\pi \left( \omega ^{2}a^{2}+c^{2}\right) ^{2}}=\frac{1}{%
4\pi a}\frac{x}{\left( x^{2}+1\right) ^{2}}\equiv I(x), 
\]
where we have introduced the dimensionless variable $x\;\equiv $ $\omega
a/c. $ Thus we get 
\[
F\left( t-t^{\prime }\right) =\frac{32\pi \U{127}h{\hskip-.2em}%
\llap{\protect\rule[1.1ex]{.325em}{.1ex}}{\hskip.2em}c}{3a^{2}}%
\int_{0}^{\infty }x\cos \left( \tau x\right) \left[ I\left( x\right) \right]
^{2}dx, 
\]
Introducing the dimensionless parameter $\tau \;\equiv $ $\left| t-t^{\prime
}\right| c/a$ and the new function 
\[
G\left( \tau \right) \equiv G(\left| t-t^{\prime }\right| c/a)=F\left(
t-t^{\prime }\right) 
\]
we get 
\begin{equation}
G\left( \tau \right) =\frac{2\U{127}h{\hskip-.2em}\llap{\protect%
\rule[1.1ex]{.325em}{.1ex}}{\hskip.2em}c}{3\pi a^{4}}\int_{0}^{\infty }\frac{%
x^{3}\cos \left( \tau x\right) dx}{\left( x^{2}+1\right) ^{4}},  \label{G}
\end{equation}
It is possible to get an analytical expression for $G\left( \tau \right) $
(an integration by parts may transform it in the integral n${{}^{o}}$
3.773/1 of Gradshteyn\cite{grad}) but we shall just obtain the behaviour for
large and small $\tau .$

For $\tau <<1,$ that is $\left| t-t^{\prime }\right| <<a/c,$ we may
approximate 
\[
\cos \left( \tau x\right) \simeq 1-\frac{1}{2}\tau ^{2}x^{2}, 
\]
and we obtain, to order $\tau ^{2},$ (with the change x$^{2}=z)$%
\[
G\left( \tau \right) \simeq \frac{\U{127}h{\hskip-.2em}\llap{\protect%
\rule[1.1ex]{.325em}{.1ex}}{\hskip.2em}c}{3\pi a^{4}}\int_{0}^{\infty }\frac{%
z\left( 1-\frac{1}{2}\tau ^{2}z\right) dz}{\left( z+1\right) ^{4}}=\frac{%
\U{127}h{\hskip-.2em}\llap{\protect\rule[1.1ex]{.325em}{.1ex}}{\hskip.2em}c}{%
18\pi a^{4}}\left( 1-\tau ^{2}\right) . 
\]

For $\tau >>1$ the main contribution to the integral will come from values
of x close to the minimum of the quantity $y\equiv x^{-3}\left(
x^{2}+1\right) ^{4},$ that is $x=\sqrt{3/5}.$ Making the change of variable $%
x=u+\sqrt{3/5}$ we may approximate, neglecting terms of order u$^{4}$ and
higher, 
\[
y\left( u\right) \simeq \frac{8^{4}}{3.5^{3}}\sqrt{\frac{5}{3}}\left( 1+%
\frac{25}{8}u^{2}\right) . 
\]
Thus we get, extending the integral in u to -$\infty $ because only values
of x close to $\sqrt{3/5}$ will contribute, 
\begin{eqnarray*}
G\left( \tau \right) &\simeq &\frac{5}{256}\sqrt{\frac{3}{5}}\frac{\U{127}h{%
\hskip-.2em}\llap{\protect\rule[1.1ex]{.325em}{.1ex}}{\hskip.2em}c}{\pi a^{4}%
}\int_{-\infty }^{\infty }\frac{\cos \left[ \left( u+\sqrt{3/5}\right) \tau
\right] du}{\left( u^{2}+8/25\right) } \\
&=&\frac{5}{256}\sqrt{\frac{3}{5}}\frac{\U{127}h{\hskip-.2em}%
\llap{\protect\rule[1.1ex]{.325em}{.1ex}}{\hskip.2em}c}{\pi a^{4}}\cos
\left( \sqrt{\frac{3}{5}}\tau \right) \int_{-\infty }^{\infty }\frac{\cos
\left( \tau u\right) du}{\left( u^{2}+8/25\right) } \\
&=&\frac{25}{512}\sqrt{\frac{3}{10}}\frac{\U{127}h{\hskip-.2em}%
\llap{\protect\rule[1.1ex]{.325em}{.1ex}}{\hskip.2em}c}{a^{4}}\exp \left( -%
\frac{2\sqrt{2}}{5}\tau \right) \cos \left( \sqrt{\frac{3}{5}}\tau \right)
.\ 
\end{eqnarray*}

\subsection{Evaluation of $\sigma $}

From eq.$\left( \ref{point}\right) $ it is natural to assume 
\[
\sigma ^{2}=\frac{c^{2}}{8\pi ^{2}}\int_{-\infty }^{\infty }F(t)^{2}dt. 
\]
The integral may be obtained from eq.$\left( \ref{G}\right) $ performing the 
$\tau $-integration first. Taking into account that 
\[
\int_{-\infty }^{\infty }\cos (x\tau )\cos \left( x^{\prime }\tau \right)
dt=\pi \left[ \delta \left( x-x^{\prime }\right) +\delta \left( x+x^{\prime
}\right) \right] , 
\]
we get 
\[
\sigma ^{2}=\frac{126\pi \U{127}h{\hskip-.2em}\llap{\protect%
\rule[1.1ex]{.325em}{.1ex}}{\hskip.2em}^{2}c^{3}}{9a^{3}}\int_{0}^{\infty
}x^{2}I(x)^{4}dx=\frac{\U{127}h{\hskip-.2em}\llap{\protect%
\rule[1.1ex]{.325em}{.1ex}}{\hskip.2em}^{2}c^{3}}{18\pi ^{3}a^{7}}%
\int_{0}^{\infty }\frac{x^{6}}{\left( x^{2}+1\right) ^{8}}dx, 
\]
giving 
\[
\sigma =\frac{2\sqrt{5}\pi \U{127}h{\hskip-.2em}\llap{\protect%
\rule[1.1ex]{.325em}{.1ex}}{\hskip.2em}c^{3/2}}{3\left( 4\pi a\right) ^{7/2}}%
\simeq 2.12\times 10^{-4}\frac{\U{127}h{\hskip-.2em}\llap{\protect%
\rule[1.1ex]{.325em}{.1ex}}{\hskip.2em}c^{3/2}}{a^{7/2}} 
\]


\begin{thebibliography}{99}
\bibitem{lamb}  Prof. Lamb has expressed the opinion that his experiment
provided an empirical proof that the vacuum is not empty. This is related to
his statement that the photon is the quantum of the electromagnetic field,
but not a particle, see W. E. Lamb Jr., \textit{Appl. Phys.} \textbf{B 60},
77 (1995). The point is important for the present paper because it stresses
the difference between photon counting and detection of true particles, like

electrons or atoms. Indeed it is not difficult to detect atoms with
practically 100\% efficiency and no noise, but this is not the case with
photons.
\end{thebibliography}
\end{document}